\documentclass[11pt]{article}
\usepackage{fullpage}
\usepackage{amsmath}
\usepackage{amsfonts}
\usepackage{amssymb}
\usepackage[dvips]{epsfig}

\def\##1{\underline{#1}}
\def\=#1{\underline{\underline{#1}}}

\def\+#1{\underline{\bf #1}}
\def\*#1{\underline{\underline{\bf #1}}}

\def\.{\mbox{ \tiny{$^\bullet$} }}
\def\I{\*I}

\def\eps{\epsilon}
\def\epso{\epsilon_{\scriptscriptstyle 0}}
\def\muo{\mu_{\scriptscriptstyle 0}}
\def\ko{k_{\scriptscriptstyle 0}}

\def\l#1{\label{#1}}
\def\r#1{(\ref{#1})}

\def\le{\left(}
\def\ri{\right)}
\def\les{\left[}
\def\ris{\right]}
\def\lec{\left\{}
\def\ric{\right\}}

\begin{document}

\begin{center}
\Large{\bf {\LARGE On convergence of the extended
strong--property--fluctuation theory for bianisotropic homogenized
composites }}

\normalsize \vspace{6mm}

Jiajia Cui\footnote{email:  s0457353@sms.ed.ac.uk} and  Tom G.
Mackay\footnote{  email: T.Mackay@ed.ac.uk}

\vspace{4mm}

\noindent{ \emph{School of Mathematics,  University of Edinburgh,\\
 Edinburgh EH9 3JZ,
United Kingdom.} }

\vspace{12mm}

{\bf Abstract}\end{center}
 The strong--property--fluctuation theory
(SPFT) provides a sophisticated means of estimating the effective
constitutive parameters of a homogenized composite material (HCM),
which takes account of the statistical distribution of the component
particles. We present an  extended version of the third--order SPFT
in which the component particles are represented as depolarization
regions of nonzero volume. Numerical results are provided for a
bianisotropic homogenization  scenario wherein the HCM is a Faraday
chiral medium. Thereby, convergence of the extended SPFT at the
second--order level of approximation is demonstrated within the
long--wavelength regime.

\vspace{8mm}





\section{Introduction}

Suppose that two (or more) homogeneous materials are blended
together to form a composite. If the length scale of inhomogeneities
in the composite is $\beta$, then the composite is  inhomogeneous at
  wavelengths $\lesssim \beta$, but homogeneous at
wavelengths $\gg \beta$ (Lakhtakia, 1996). The homogenized composite
material (HCM) which arises in the long--wavelength regime may
exhibit properties that are not exhibited by its component
materials, or at least not exhibited to the same extent. Thus, HCMs
represent prime examples of metamaterials (Walser, 2003). Interest
in complex metamaterials, and HCM--based complex metamaterials in
particular, has escalated in recent years, which serves to highlight
the need for accurate formalisms to estimate the constitutive
parameters of complex HCMs (Mackay, 2005).

The strong--property--fluctuation theory (SPFT) provides a
sophisticated  basis for estimating the constitutive parameters of
HCMs (Tsang \& Kong, 1981) which has distinct advantages over
conventional formalisms such as those named after Maxwell Garnett
and Bruggeman (Lakhtakia, 1996; Ward, 1995). This is achieved
through accommodating higher--order descriptions of the
distributional statistics of the component materials which are
brought together to form the HCM. Indeed, in principle, the SPFT can
accommodate spatial correlation functions of arbitrarily high order.
However, in practice, the SPFT is usually implemented at the
second--order level of approximation wherein  a two--point
covariance function and its associated correlation length
characterize the distributional statistics of the component
materials. Versions of the second--order SPFT have been developed
for isotropic (Tsang \& Kong, 1981; Michel \& Lakhtakia, 1995),
anisotropic (Genchev, 1992; Zhuck, 1994) and bianisotropic (Mackay,
Lakhtakia, \& Weiglhofer, 2000) linear HCMs, as well as for certain
nonlinear HCMs (Lakhtakia, 2001; Mackay, Lakhtakia, \& Weiglhofer,
2003; Mackay, 2003). The third--order SPFT has also been established
for bianisotropic HCMs which are weakly nonlinear. Convergence at
the second--order level of approximation has been established for
isotropic chiral mediums, and also more generally for  bianisotropic
mediums which are weakly anisotropic (Mackay, Lakhtakia, \&
Weiglhofer, 2001a).

Commonly, in homogenization formalisms~---~including the SPFT~---~
the electromagnetic responses of the component material particles
are represented by depolarization dyadics. Often the depolarization
dyadics are taken to correspond to  vanishingly small regions;
accordingly, the spatial extent of the component material particles
is neglected (Michel, 1997; Michel \& Weiglhofer, 1997). An extended
version of the second--order SPFT has recently been established
which allows for depolarization regions of nonzero volume (Cui \&
Mackay, 2007a).
 Numerical studies
based on the extended SPFT have demonstrated that the depolarization
contribution associated with  regions of nonzero volume can have
significant effects on  estimates of the HCM constitutive parameters
(Cui \& Mackay, 2007a,b). We note that similarly extended versions
of the Maxwell Garnett (Lakhtakia \& Shanker, 1993; Shanker \&
Lakhtakia, 1993a,b) and Bruggeman (Prinkey, Lakhtakia, \& Shanker,
1994; Shanker, 1996) homogenization formalisms have also been
established, and the spatial extent of the component material
particles has been emphasized in other homogenization studies too
(Doyle, 1989; Dungey \& Bohren, 1991). However, these  studies did
not incorporate higher--order statistical details concerning the
distribution of the component material particles, nor did they
consider the most general linear scenario represented by
bianisotropic HCMs.

 In this paper we present the extended third--order SPFT for
 biansotropic HCMs which are weakly anisotropic, and investigate
 the convergence of the extended SPFT.

The following notation is used: Vector quantities are underlined.
Double underlining  and normal (bold) face signifies a 3$\times$3
(6$\times$6) dyadic.
 The inverse  of
a dyadic $\*M$ is denoted by $\*M^{-1}$. The 3$\times$3 (6$\times$6)
identity dyadic
 is represented by  $\,\=I\,$ ($\,\*I\,$).
All field--related quantities are implicitly functions of the
angular frequency $\omega$.
 The permittivity and
permeability of free space are denoted as $\epso$ and $\muo$,
respectively; the free--space wavenumber is $\ko = \omega \sqrt{
\epso \muo }\,$.  The real and imaginary parts of $z \in \mathbb{C}$
are represented by $\mbox{Re} \,z$ and $\mbox{Im} \,z$,
respectively. A compact representation of the constitutive
parameters for the homogeneous bianisotropic material specified by
the Tellegen constitutive relations
\begin{equation}
\left.
\begin{array}{l}
\#D(\#r) = \=\eps \. \#E(\#r) + \=\xi \. \#H(\#r) \\
\#B(\#r) = \=\zeta \. \#E(\#r) + \=\mu \. \#H(\#r)
\end{array}
\right\}
\end{equation}
is provided by  the 6$\times$6 constitutive dyadic
\begin{equation}
\*K_{\,} = \les \begin{array}{cc} \=\eps_{\, } & \=\xi_{\, }
\\ \=\zeta_{\,} & \=\mu_{\,} \end{array} \ris.
\end{equation}
Herein, $\=\eps_{\, }$ and $\=\mu_{\, }$ are the  3$\times$3
permittivity and  permeability dyadics , respectively, while
$\=\xi_{\, }$ and $\=\zeta_{\, }$ are the 3$\times$3
magneto\-electric dyadics. Subscripts on $\*K$  identify the
particular material that $\*K$ describes.

\section{Strong--property--fluctuation theory} \l{SPFT_sec}

\subsection{Component materials}

We consider the homogenization of two distinct material phases:
phase $a$ and phase $b$, both of which consist of spherical
particles of average radius $\eta$. All space is taken to be
partitioned into the disjoint regions $V_a$ and $V_b$ that contain
the phases $a$ and $b$, respectively.
 The  phases $a$ and $b$
are randomly distributed, as specified by the characteristic
functions
\begin{equation}
\Phi_\ell(\#r) = \left\{ \begin{array}{ll} 1,\, &  \qquad \#r \in
V_\ell\,\\ & \qquad \qquad \qquad, \qquad  \qquad (\ell =a,b). \\
  0, \,&  \qquad \#r \not\in V_\ell\, \end{array} \right. \l{theta}
\end{equation}
In particular, within the SPFT statistical  moments of $\Phi_\ell$
are utilized to characterize the component phase distributions. The
volume fraction of phase $\ell$ is given by the first moment; i.e.,
\begin{equation}
 \langle \, \Phi_\ell(\#r) \, \rangle = f_\ell , \qquad \qquad
(\ell=a,b);
\end{equation}
and we have $f_a + f_b = 1$. The physically--motivated step function
(Tsang, Kong, \& Newton, 1982)
\begin{equation}
\langle \, \Phi_\ell(\#r) \, \Phi_\ell(\#r') \, \rangle = \left\{
\begin{array}{ll}  f_\ell,  &  \qquad  |\,\#r - \#r'\,|
\leq L\\  &\qquad \qquad \qquad \qquad, \qquad  \qquad \le \ell = a,b \ri \\
f^2_\ell, & \qquad
 |\,\#r - \#r'\,|
> L \end{array} \right. \l{cov2}
\end{equation}
is often adopted  as the second moment for the second--order SPFT.
The correlation length $L$ is required to be much smaller than the
electromagnetic wavelength(s) but larger than the size of the
component phase particles. It is worth noting that the second--order
SPFT estimates of the HCM constitutive parameters have been found to
be largely insensitive to the particular form of the second moment
(Mackay, Lakhtakia, \& Weiglhofer, 2001b). In keeping with \r{cov2},
the third--order SPFT has been established for the  third moment
(Mackay, Lakhtakia, \& Weiglhofer, 2001a)
\begin{equation}
\langle \, \Phi_\ell(\#r) \, \Phi_\ell(\#r') \, \Phi_\ell(\#r'')
\,\rangle = \left\{
\begin{array}{ll} f^3_\ell,  &  \qquad  \; \mbox{min}
 \{L_{12}, L_{13}, L_{23} \}
> L\\ & \qquad \qquad \qquad \qquad  \\
 f_\ell, & \qquad
 \; \mbox{max}
 \{L_{12}, L_{13}, L_{23} \}
\leq L
\\ & \qquad \qquad \qquad \qquad \qquad \qquad \qquad, \qquad \qquad \le \ell = a,b \ri,  \\
 \frac{1}{3}\,(f_\ell + 2 f^3_\ell), & \qquad
\mbox{one of} \; L_{12}, L_{13}, L_{23} \leq L
\\ & \qquad \qquad \qquad \qquad  \\
\frac{1}{3}\,(2 f_\ell +  f^3_\ell), &  \qquad \mbox{two of} \;
L_{12}, L_{13}, L_{23} \leq L
\end{array} \right. \l{cov3}
\end{equation}
where
\begin{equation}
L_{12} = | \#r - \#r' |\,, \qquad L_{13} = | \#r - \#r'' |\,, \qquad
L_{23} = | \#r' - \#r'' |\,.
\end{equation}

The component material phases $a$ and $b$ are homogeneous materials,
characterized by the 6$\times$6 constitutive dyadics $\*K_{\,a}$ and
$\*K_{\,b}$, respectively.

\subsection{Homogenized composite material}

 The constitutive dyadic of the HCM, as
estimated by the $n$th--order SPFT, is given by (Mackay, Lakhtakia,
\& Weiglhofer, 2000)
\begin{equation}
 \*K_{\,HCM}^{[n]}  =
 \*K_{\,cm}  - \frac{1}{i \omega}  \les \,\I +
\mbox{\boldmath$\={\Sigma}$}^{[n]}(\eta, L)  \. \*D (\eta)
\,\ris^{-1} \. \mbox{\boldmath$\={\Sigma}$}^{[n]} (\eta, L) .
\l{KDy0}
\end{equation}
Herein, the constitutive dyadic $\*K_{\,cm} $ characterizes a
comparison medium whose constitutive parameters are provided by the
Bruggeman homogenization formalism.

 The depolarization dyadic
\begin{equation}
\*D (\eta) =    \*D^{ 0} + \*D^{>0} (\eta)  \l{D_eta_def}
\end{equation}
has two parts (Cui \& Mackay, 2007a): $ \*D^{ 0}$ represents the
contribution to the depolarization arising from  component particles
in the limit $\eta \rightarrow 0$, whereas $\*D^{>0} (\eta)$
represents the depolarization contribution arising from the nonzero
volume of the component particles. Often in homogenization studies
the $\*D^{>0} (\eta)$ contribution is neglected, but recent studies
have highlighted the significance of this contribution, particularly
in the context of scattering losses (Cui \& Mackay, 2007a,b). The
conventional SPFT incorporates $\*D^{0}$ only as the depolarization
dyadic, whereas the extended SPFT accommodates  both $\*D^{ 0}$ and
$\*D^{>0} (\eta)$. The mathematical expressions for $\*D^{ 0}$ and
$\*D^{>0} (\eta)$ are complicated, especially for bianisotropic
HCMs, but integral representations are available which can be
straightforwardly evaluated using standard numerical techniques
(Press, Flannery, Teukolsky, \& Vetterling, 1992). These integral
representations are provided in the Appendix.

The mass operator term $\mbox{\boldmath$\={\Sigma}$}^{[n]} (\eta,
L)$ in \r{KDy0} vanishes for the zeroth--  and  first--order
versions of the SPFT (Tsang \& Kong, 1981; Mackay, Lakhtakia, \&
Weiglhofer, 2000); i.e.,
\begin{equation}
\mbox{\boldmath$\={\Sigma}$}^{[0]} =
\mbox{\boldmath$\={\Sigma}$}^{[1]} = \*0.
\end{equation}
By implementing the two--point covariance function \r{cov2}, the
second--order mass operator term is given by (Mackay, Lakhtakia, \&
Weiglhofer, 2000)
\begin{equation}
\mbox{\boldmath$\={\Sigma}$}^{[2]} (\eta, L) = f_a f_b \les
\,\mbox{\boldmath$\=\chi$}_{\,a} (\eta) -
\mbox{\boldmath$\=\chi$}_{\,b} (\eta)
 \ris \.   \*D^{>0} (L)
 \. \les \,\mbox{\boldmath$\=\chi$}_{\,a} (\eta) -
\mbox{\boldmath$\=\chi$}_{\,b} (\eta)  \ris, \l{spv}
\end{equation}
with the polarizability density dyadics
\begin{eqnarray}
 \mbox{\boldmath$\=\chi$}_{\,\ell} (\eta)  & =& - i \omega
\le\,\*K_{\,\ell}  - \*K_{\,cm}  \,\ri\.\les \,\*I +  i \omega \*D
(\eta) \. \le\, \*K_{\,\ell}  - \*K_{\,cm}  \,\ri \ris^{-1},
 \qquad  (\ell = a,b).
\end{eqnarray}
The three--point covariance function \r{cov3} yields  the
third--order mass operator term (Mackay, Lakhtakia, \& Weiglhofer,
2001a)
\begin{eqnarray}
\mbox{\boldmath$\={\Sigma}$}^{[3]} (\eta, L) &=&
\mbox{\boldmath$\={\Sigma}$}^{[2]} (\eta, L) + \frac{f_a (1 -
2f_a)}{3 (1-f_a)^2} \, \mbox{\boldmath$\=\chi$}_{\, a} (\eta)
\.\Big[ \,
\*V (\eta) \. \mbox{\boldmath$\=\chi$}_{\, a} (\eta)  \.  \*D^{>0} (L) \nonumber \\
&& \,+\,
  \*D^{>0} (L) \. \mbox{\boldmath$\=\chi$}_{\, a} (\eta)  \. \*V  (\eta) \,+ \,  \*D^{>0} (L) \.
\mbox{\boldmath$\=\chi$}_{\, a} (\eta) \.  \*D^{>0} (L) \,\Big] \.
\mbox{\boldmath$\=\chi$}_{\, a} (\eta),
\end{eqnarray}
where
\begin{equation}
\*V  (\eta) = \frac{1}{i \omega}\, \*K^{-1}_{\,cm} - \*D (\eta).
\end{equation}

\section{Numerical studies} \l{num_studies}

We now apply  the  extended third--order SPFT presented in
\S\ref{SPFT_sec} to a specific bianisotropic homogenization
scenario. As an illustrative
  example,
 let us
consider the homogenization of (i) a magnetically--biased ferrite
medium described by the constitutive dyadic (Lax \& Button, 1962;
Collin, 1966)
\begin{equation}
\*K_{\,a} = \delta \les \begin{array}{cc} \epso \, \eps_a \, \=I &
\=0
\\ \vspace{-8pt} & \\ \=0 & \muo \,
\le
\begin{array}{ccc} \mu^x_a & i \mu^g_a & 0 \\
\vspace{-12pt}
&& \\
-i \mu^g_a & \mu^x_a & 0 \\
\vspace{-12pt}
&& \\
0 & 0 & \mu^z_a
\end{array}
\ri
 \end{array} \ris \l{mbf}
\end{equation}
and (ii) an isotropic chiral medium described by the constitutive
dyadic (Lakhtakia, 1994)
\begin{equation}
\*K_{\,b} = \les \begin{array}{cc} \epso \eps_b \, \=I & i
\sqrt{\epso \muo} \, \xi_b \, \=I
\\ \vspace{-8pt} & \\ - i
\sqrt{\epso \muo} \,\xi_b \, \=I &  \muo \mu_b \, \=I\end{array}
\ris. \l{iso_chiral}
\end{equation}
The parameter $\delta$ in \r{mbf} provides a means of varying the
constitutive contrast between the component material phases. The
constitutive relations of the resulting HCM~---~which is known as a
Faraday chiral medium~---~are rigorously established (Engheta,
Jaggard, \& Kowarz, 1992; Weiglhofer \& Lakhtakia, 1998). The
constitutive dyadic of the  HCM, as estimated by the $n$th order
SPFT,  has the general form
\begin{equation}
\*K^{[n]}_{\,HCM} = \les \begin{array}{cc} \epso \,  \le
\begin{array}{ccc} \eps^x_{HCM} & i \eps^g_{HCM} & 0 \\
\vspace{-12pt}
&& \\
-i \eps^g_{HCM} & \eps^x_{HCM} & 0 \\
\vspace{-12pt}
&& \\
0 & 0 & \eps^z_{HCM}
\end{array}
\ri
 & i \sqrt{\epso
\muo} \,
  \le
\begin{array}{ccc} \xi^x_{HCM} & i \xi^g_{HCM} & 0 \\
\vspace{-12pt}
&& \\
-i \xi^g_{HCM} & \xi^x_{HCM} & 0 \\
\vspace{-12pt}
&& \\
0 & 0 & \xi^z_{HCM}
\end{array}
\ri
\\ \vspace{-8pt} & \\ - i \sqrt{\epso
\muo} \,
  \le
\begin{array}{ccc} \xi^x_{HCM} & i \xi^g_{HCM} & 0 \\
\vspace{-12pt}
&& \\
-i \xi^g_{HCM} & \xi^x_{HCM} & 0 \\
\vspace{-12pt}
&& \\
0 & 0 & \xi^z_{HCM}
\end{array}
\ri & \muo \,
  \le
\begin{array}{ccc} \mu^x_{HCM} & i \mu^g_{HCM} & 0 \\
\vspace{-12pt}
&& \\
-i \mu^g_{HCM} & \mu^x_{HCM} & 0 \\
\vspace{-12pt}
&& \\
0 & 0 & \mu^z_{HCM}
\end{array}
\ri
\end{array} \ris. \l{K_HCM_fcm1}
\end{equation}
 An HCM
of the same form also arises from the homogenization of a
magnetically--biased plasma and an isotropic chiral medium
(Weiglhofer, Lakhtakia, \& Michel, 1998; Weiglhofer \& Mackay,
2000). Bearing in mind that the third--order SPFT is established
only for bianisotropic mediums which are weakly anisotropic (Mackay,
Lakhtakia, \& Weiglhofer, 2001a), we select the following
representative values for the constitutive parameters of the
component material phases: $\eps_a = 1.2 + i 0.02$, $\mu^x_a = 3.5 +
i 0.08$, $\mu^g_a = 0.7 + i 0.005$, $\mu^z_a = 3.0 + i 0.06$;
$\eps_b = 2.5 + i 0.1$, $\xi_b = 1 + i 0.07$ and $\mu_b = 1.75 + i
0.09$. Results which are qualitatively similar to those presented
here were observed~---~in further studies not reported here~---~when
different values were selected for the constitutive parameters of
the component materials.

In the following numerical studies,  the correlation length $L$ is
fixed\footnote{This applies to
 the second--order and third--order SPFT calculations; $L$ does not feature in the
zeroth--order SPFT.}  for each value of $\delta$ considered, while
 the  particle size parameter $\eta$ varies from 0 to $L/2$.
  In order to conform to the long--wavelength
regime under which the SPFT estimates of the HCM parameters are
derived, the value of $L$ is selected such that the scalar  $Q \ll
1$, where
\begin{equation}
Q = \frac{\mbox{max}\, \lec | \gamma_1 |,  \,| \gamma_2 |,\, |
\gamma_3 |,\, | \gamma_4 |\ric}{2 \pi}\, L,
\end{equation}
with $\lec \gamma_i \, | \, i=1,\ldots,4 \ric$  being the four
independent wavenumbers supported by the HCM. For simplicity, we
choose the wavenumbers associated with propagation along the
Cartesian $z$ axis (Mackay \& Lakhtakia, 2004); i.e.,
\begin{equation}
\left.
\begin{array}{l}
\gamma_1 = \ko \le \sqrt{\eps^x_{HCM} +
\eps^g_{HCM}}\sqrt{\mu^x_{HCM} +
\mu^g_{HCM}} -\xi^x_{HCM} - \xi^g_{HCM} \ri \vspace{4pt} \\
\gamma_2 = \ko \le - \sqrt{\eps^x_{HCM} +
\eps^g_{HCM}}\sqrt{\mu^x_{HCM} + \mu^g_{HCM}} -\xi^x_{HCM} -
\xi^g_{HCM} \ri
 \vspace{4pt} \\
\gamma_3 = \ko \le  \sqrt{\eps^x_{HCM} -
\eps^g_{HCM}}\sqrt{\mu^x_{HCM} - \mu^g_{HCM}} +\xi^x_{HCM} -
\xi^g_{HCM} \ri  \vspace{4pt} \\
\gamma_4 = \ko \le - \sqrt{\eps^x_{HCM} -
\eps^g_{HCM}}\sqrt{\mu^x_{HCM} - \mu^g_{HCM}} +\xi^x_{HCM} -
\xi^g_{HCM} \ri
\end{array}
\right\}.
\end{equation}
 All  numerical calculations were carried out using an angular
frequency $\omega = 2 \pi \times 10^{10}$ $\mbox{rad}\,
\mbox{s}^{-1}$ with the volume fraction fixed at
  $f_a = 0.3$.

\begin{figure}[!ht]
\centering \psfull \epsfig{file=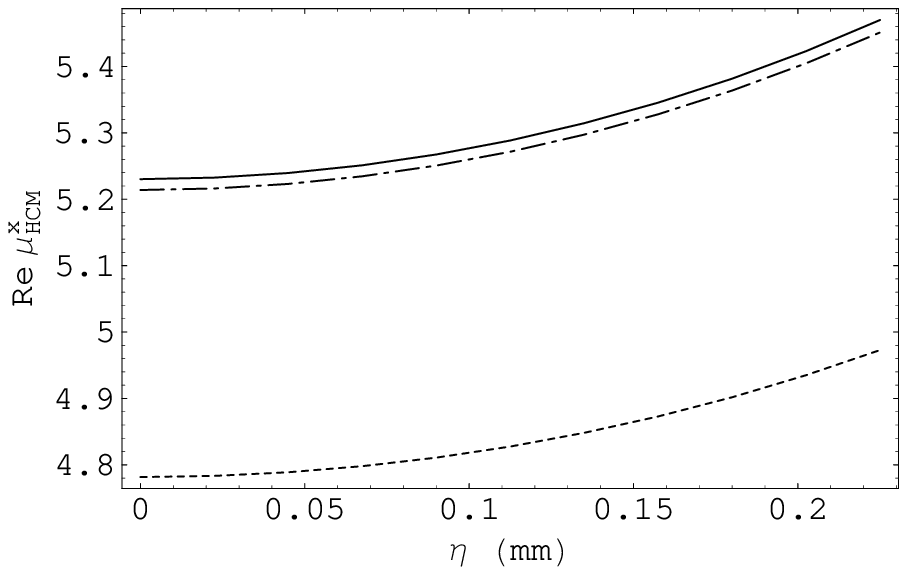,width=3.2in} \hfill
  \epsfig{file=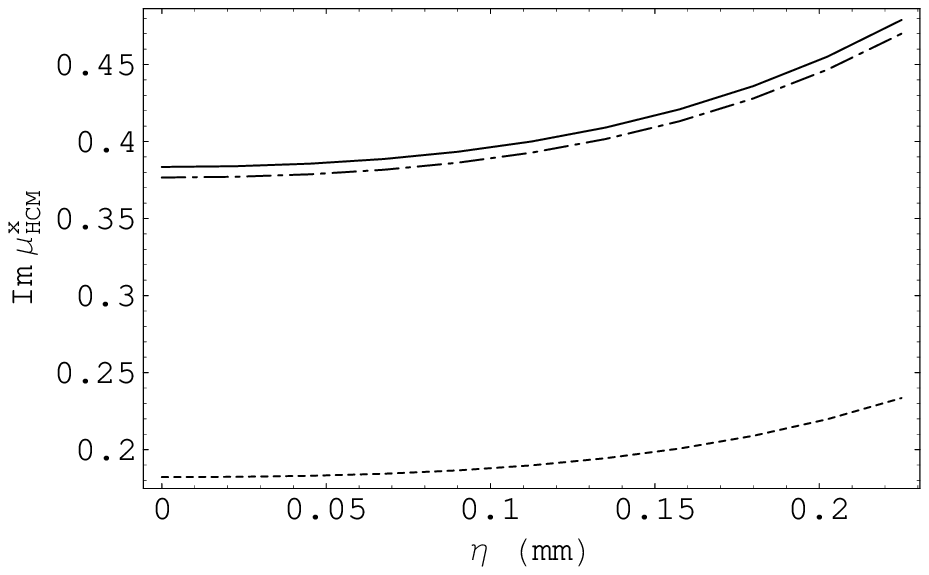,width=3.2in} \\
   \epsfig{file=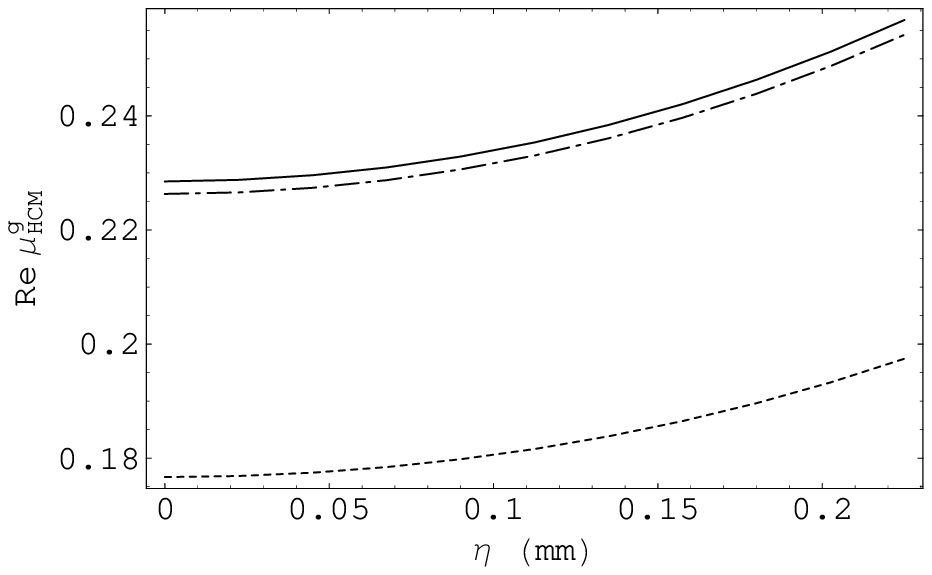,width=3.2in} \hfill
  \epsfig{file=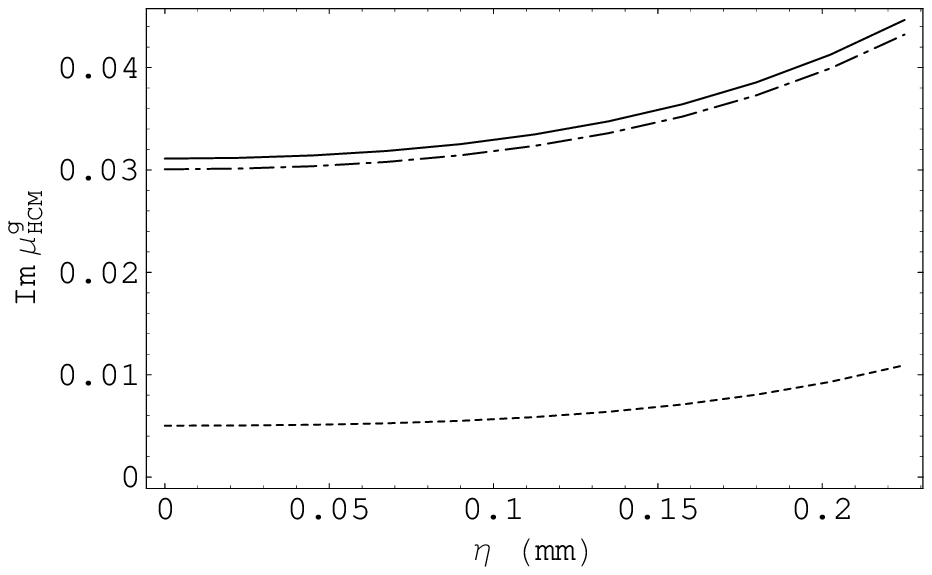,width=3.2in}\\
   \epsfig{file=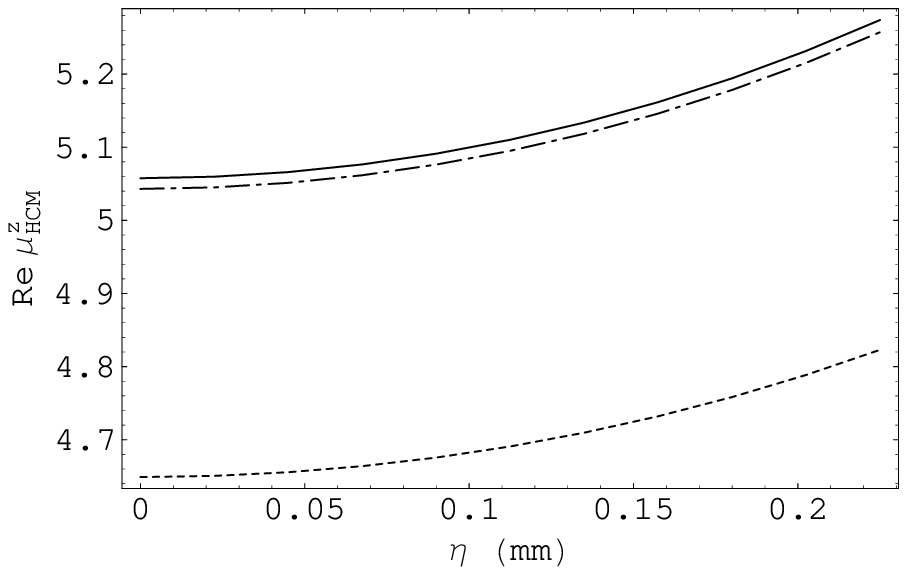,width=3.2in} \hfill
  \epsfig{file=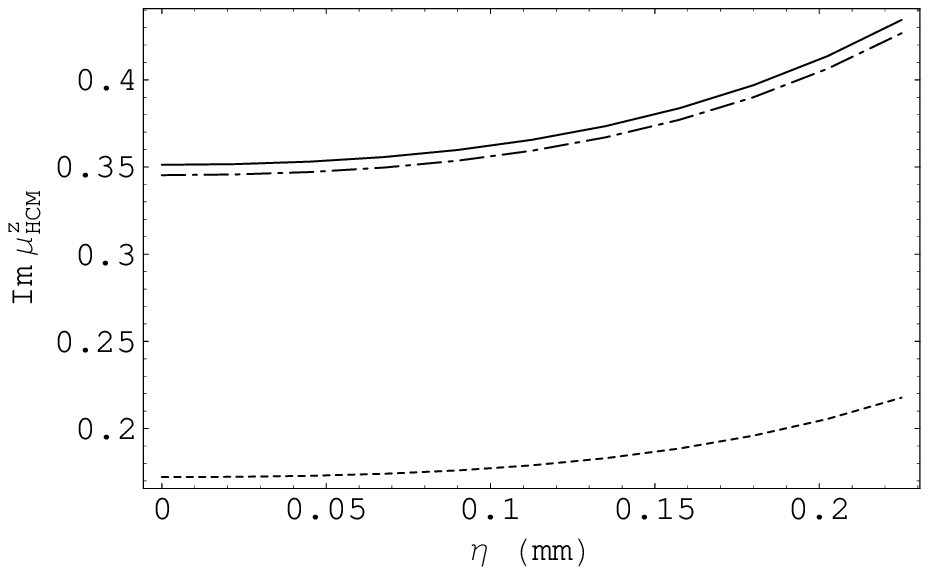,width=3.2in}
\caption{ \label{fig1} Real (left) and imaginary (right) parts of
the HCM constitutive parameters  $\mu^{x,z,g}_{HCM}$  plotted
against $ \eta $ (mm) for $\delta = 10$. Key: dashed curve is the
zeroth--order SPFT estimate; broken dashed curve is the
second--order SPFT estimate; and solid curve is the third--order
SPFT estimate. }
\end{figure}

The  zeroth--, second-- and third--order SPFT estimates of the HCM
constitutive parameters $\mu^x_{HCM}$, $\mu^g_{HCM}$ and
$\mu^z_{HCM}$ are plotted against the size parameter $\eta$ in
Figure~\ref{fig1} for the case where $\delta = 10$. Here, the
correlation length is set at $L = 0.45$ mm, in order that $Q= 0.1$
at $\eta = L/2$.  The third--order estimates of the real and
imaginary parts of $\mu^x_{HCM}$, $\mu^g_{HCM}$ and $\mu^z_{HCM}$
increase steadily as $\eta$ is increased, as do the zeroth--order
and second--order estimates, but the difference between the
third--order estimates and the second--order estimates remains very
small for all values of $\eta$. In contrast, there are plainly
significant differences between the second--order and zeroth--order
estimates. Furthermore, the difference between the zeroth-- and
second--order estimates increases in magnitude slightly as the the
size parameter $\eta$ increases. The corresponding graphs for the
permittivity and magnetoelectric constitutive parameters of the HCM
are qualitatively similar to those graphs presented in
Figure~\ref{fig1}. Accordingly, these are not displayed here.

\begin{figure}[!ht]
\centering \psfull \epsfig{file=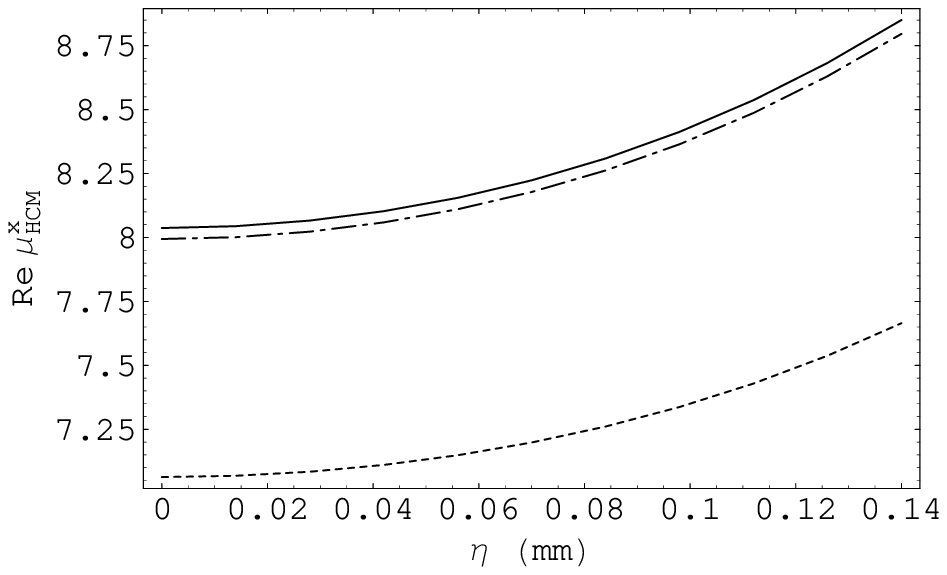,width=3.2in} \hfill
  \epsfig{file=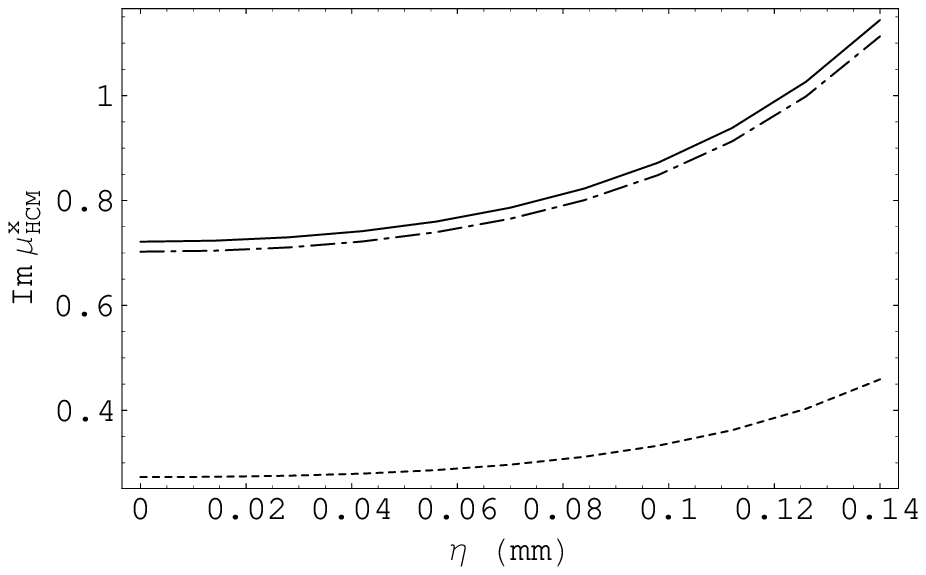,width=3.2in} \\
   \epsfig{file=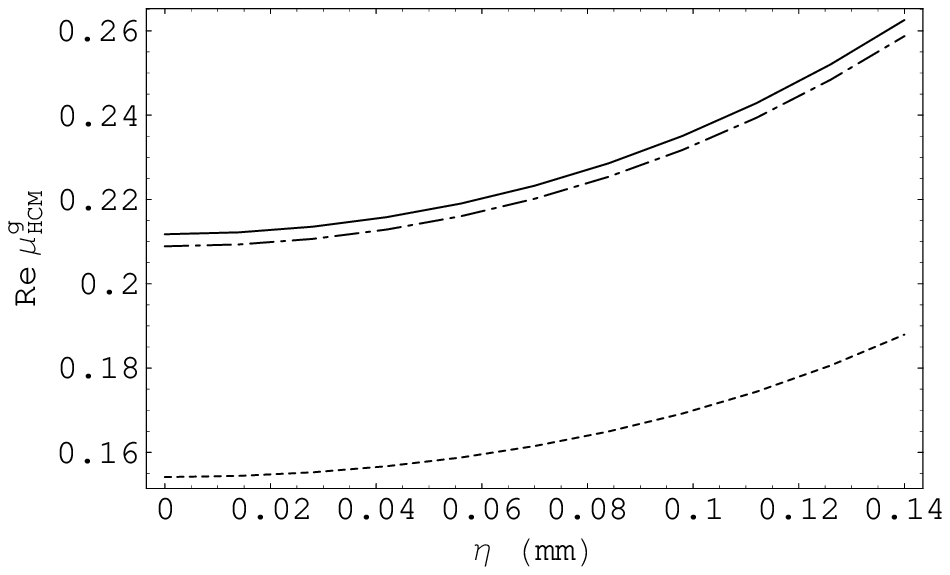,width=3.2in} \hfill
  \epsfig{file=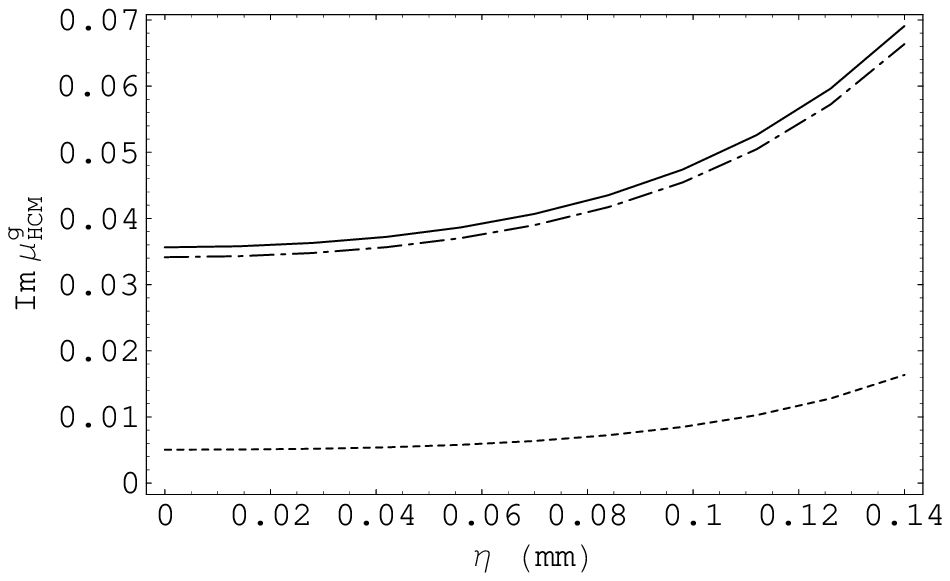,width=3.2in}\\
   \epsfig{file=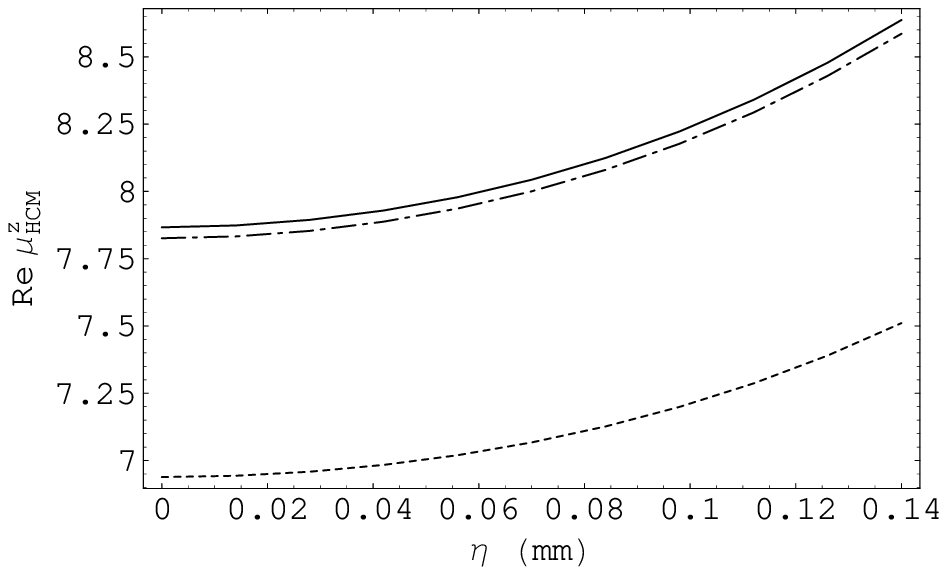,width=3.2in} \hfill
  \epsfig{file=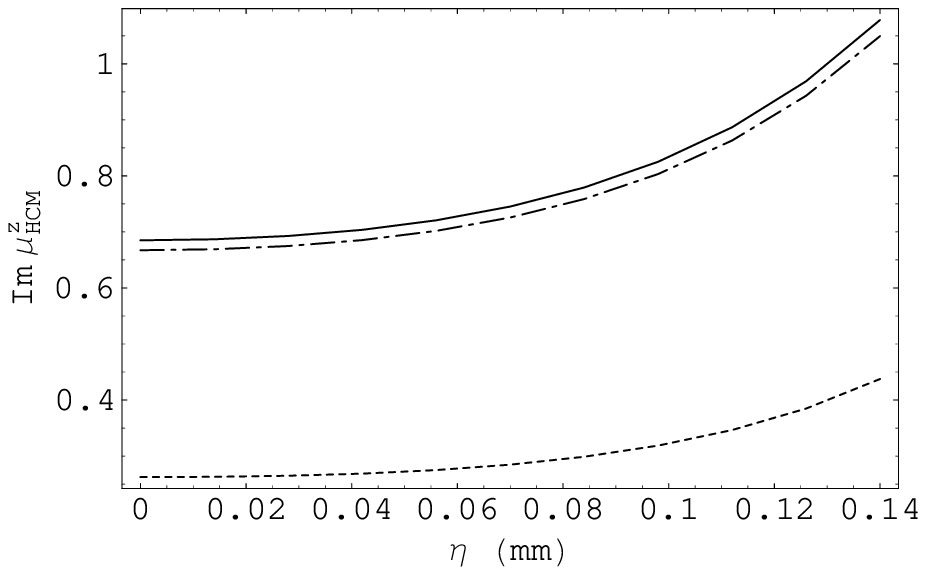,width=3.2in}
\caption{ \label{fig3} As Figure~\ref{fig1} but for $\delta = 30$. }
\end{figure}

Plots of $\mu^x_{HCM}$, $\mu^g_{HCM}$ and $\mu^z_{HCM}$ versus
$\eta$  for the case where $\delta = 30$ are shown in
Figure~\ref{fig3}. The correlation length $L = 0.28$ mm was used for
the calculations of Figure~\ref{fig3}, thereby resulting in $Q =
0.1$ at $\eta = L/2$. As is the case for $\delta = 10$, the
second--order estimates are plainly different to the zeroth--order
estimates for $\delta =  30$. The differences between second--order
and third--order estimates of the real and imaginary parts of the
HCM constitutive parameters increase slightly as $\delta$ increases,
but they remain small for all values of $\eta$. The corresponding
permittivity and magnetoelectric constitutive parameters of the HCM
exhibit trends which are qualitatively similar to those shown in
Figure~\ref{fig3} for the HCM magnetic constitutive parameters.

\section{Concluding remarks} \l{conc_remarks}

It is demonstrated by our numerical studies  in \S\ref{num_studies}
(and in further studies not presented here) that the extended SPFT
at the third--order level of approximation does not add
significantly to the HCM estimates yielded by the second--order
extended SPFT. The differences between second--order and
third--order estimates of the HCM constitutive parameters are very
small for all values of the size parameter investigated, even when
the constitutive contrast between the component materials is as
large as a factor of 30. Significant differences between the
second-- and third--order estimates arise only when (i) the
correlation length and/or size parameter become similar in magnitude
to the electromagnetic wavelength(s);
 and/or (ii) the constitutive contrast between the
component materials becomes enormous. In the case of (i) the bounds
imposed by the long--wavelength regime are exceeded, while in the
case of (ii) the contrast between  the polarizability density
dyadics $\mbox{\boldmath$\=\chi$}_{\,a}$  and
$\mbox{\boldmath$\=\chi$}_{\,b}$ is likely to become strong. In
either scenario the basic assumptions underlying the
long--wavelength SPFT are violated (Mackay, Lakhtakia, \&
Weiglhofer, 2000, 2001a). We therefore conclude that the extended
SPFT converges at the second--order level of approximation for
bianisotropic HCMs which are weakly anisotropic.

\vspace{10mm} \noindent {\bf Acknowledgement:} JC is supported by a
\emph{Scottish Power--EPSRC  Dorothy  Hodgkin Postgraduate Award}.
TGM is supported by a \emph{Royal Society of Edinburgh/Scottish
Executive Support Research Fellowship}.

\vspace{10mm}

\section*{Appendix}

The depolarization dyadic $\*D^{} (\eta)$, as specified by
\r{D_eta_def}, for a spherical inclusion of radius $\eta$, immersed
in a bianisotropic comparison medium described by the constitutive
dyadic $\*K_{\,cm}$, is derived from the dyadic Green function of
the comparison medium. The contribution associated with $\eta
\rightarrow 0$ is provided by the $\eta$--independent surface
integral (Michel, 1997; Michel \& Weiglhofer, 1997)
\begin{equation} \l{dd_def}
\*D^0 = \frac{1}{4 \pi i \omega } \int^{2 \pi}_{\phi = 0} \,
\int^{\pi}_{\theta = 0} \,\underline{\underline{\bf
B}}_{\,cm}(\hat{\#q})\;\; \sin \theta \; d\theta \; d\phi,
\end{equation}
with dyadic integrand
\begin{eqnarray}
\underline{\underline{\bf B}}_{\,cm}(\hat{\#q}) &=& \frac{1}{ b(
\theta, \phi )} \les
\begin{array}{ccc} \alpha_\mu ( \theta, \phi )\, \hat{\#q}\,\hat{\#q} && - \alpha_\zeta ( \theta, \phi ) \,
\hat{\#q}\,\hat{\#q}
\\ \vspace{-8pt} & \\
- \alpha_\xi ( \theta, \phi ) \, \hat{\#q}\,\hat{\#q} && \alpha_\eps
( \theta, \phi ) \, \hat{\#q}\,\hat{\#q}
\end{array} \ris.
\end{eqnarray}
Herein the scalars
\begin{equation}
\alpha_P ( \theta, \phi ) = \hat{\#q}\.  \={P}_{\,cm} \.\hat{\#q}\,,
\qquad \qquad (P = \eps, \zeta, \xi, \mu)
\end{equation}
and
\begin{equation}
b( \theta, \phi ) = \les \alpha_\eps ( \theta, \phi )\, \alpha_\mu (
\theta, \phi ) \ris - \les \alpha_\xi ( \theta, \phi )\,
\alpha_\zeta ( \theta, \phi )\ris,
\end{equation}
while $\hat{\#q}$ is the radial unit vector specified by the
spherical coordinates $\theta$ and $\phi$.

The depolarization contribution associated with the nonzero
inclusion volume is given by the $\eta$--dependent volume integral
(Mackay, 2004; Cui \& Mackay, 2007a)
\begin{eqnarray}
&& \*D^{>0} (\eta) =  \frac{\eta}{2 \pi^2 i \omega } \, \int_{\#q}
\frac{1}{q^2} \, \les \frac{\sin (q \eta )}{q \eta} - \cos ( q \eta)
\ris \les \underline{\underline{\bf A}}^{-1}_{\, cm}(\#q) -
\underline{\underline{\bf B}}_{\,cm}(\hat{\#q}) \ris \; d^3 \#q ,
\l{D0}
\end{eqnarray}
wherein
\begin{equation}
\underline{\underline{\bf A}}_{\, cm}(\#q)= \les
\begin{array}{cc} \=0 &  (\#q /\omega) \times \=I \\ \vspace{-8pt} & \\
-(\#q /\omega) \times \=I & \=0
\end{array} \ris +
\underline{\underline{\bf K}}_{\, cm}. \l{Gq_Ad}
\end{equation}
The depolarization integrals \r{dd_def} and \r{D0} are
straightforwardly evaluated using standard numerical techniques
(Press, Flannery, Teukolsky, \& Vetterling, 1992).

\newpage

\newpage





\end{document}